\documentclass[conference]{IEEEtran}
\IEEEoverridecommandlockouts 

\usepackage[utf8]{inputenc}
\usepackage[T1]{fontenc}
\usepackage{cite}
\usepackage{amsmath,amssymb,amsfonts}
\usepackage{graphicx}
\usepackage{textcomp}
\usepackage{xcolor}
\usepackage{hyperref}
\usepackage{booktabs}
\usepackage{array}
\usepackage{url}
\usepackage{xurl}

\usepackage[russian, english]{babel}

\hypersetup{colorlinks=true,linkcolor=blue,citecolor=blue,urlcolor=blue}

\begin{document}

\title{Inside Crypter-as-a-Service: An Ecosystem Analysis of the exploit.in Underground Forum \\ Research Talks
}
\author{
\IEEEauthorblockN{
Mathieu Jeannot, Jean-Yves Marion, Manon Pamar, Maira Nassau, Pierre Marty, Romain Guittienne
}
\IEEEauthorblockA{
Université de Lorraine, CNRS, LORIA\\
Nancy, France\\
jean-yves.marion@loria.fr
\thanks{\textbf{Paper under submission. Please do not distribute.}}
}
}

\maketitle

\begin{abstract}
Crypter-as-a-Service (CraaS) has become a key enabling layer of the contemporary malware economy by providing on-demand evasion capabilities through underground service markets. In this paper, we present a longitudinal characterization of the CraaS ecosystem on \texttt{exploit.in}, a major Russian-language cybercrime forum with a presence on both the clear web and the dark web.

From a collection of approximately 1,000,000 posts, we combine keyword filtering, LLM-assisted annotation, and manual validation to extract a corpus of 491 threads and 2,949 posts spanning January 2020 to August 2025.

Our analysis shows that crypters on \texttt{exploit.in} are not merely sold as static tools, but as continuously maintained operational services whose value depends on recurring stub renewal --- sometimes on a daily basis --- sustained antivirus evasion, and trust-based delivery. We develop a taxonomy of five seller types and four buyer profiles, and map the buyer--seller correspondences that structure market transactions. We further document pricing models ranging from low-cost per-build Telegram bot services to high-end custom development and salaried recruitment.

Using social-network analysis, we find that the market is hierarchically structured around a small core of highly central actors, many of whom appear to function as trust brokers or other influential intermediaries, while its stability relies on a broader trust and governance infrastructure including escrow, guarantors, reputation systems, and security deposits.

Finally, we discuss differences between the CraaS model observed on \texttt{exploit.in} and that reported on \texttt{HackForums}. Although both forums share similar service logics, our corpus suggests that \texttt{exploit.in} exhibits a more professionalized and service-oriented CraaS configuration.
\end{abstract}


\section{Introduction}
A \emph{crypter} is a software tool that transforms a malicious binary through a combination of encryption and obfuscation, producing a functionally equivalent payload capable of evading antivirus~(AV) detection --- a state often referred to in underground communities as \emph{Fully UnDetectable} (FUD). Architecturally, a crypter typically consists of two components: the \emph{builder}, which encrypts and obfuscates the original binary and binds it to an executable stub, and the \emph{stub}, which decrypts and executes the payload directly in memory with explicit evasion objectives~\cite{Muralidharan2022}.

Beyond this technical architecture, crypters have evolved into a highly commoditized segment of the cybercrime underground. Through the \emph{Crypter-as-a-Service}~(CraaS)\footnote{We use \emph{CraaS} (Crypter-as-a-Service) to avoid confusion with other ``CaaS'' acronyms.} model, specialized vendors offer subscription-based access, guaranteed freshness windows, automated stub updates, dedicated support channels, and replacement or refund policies when advertised detection-free periods are not met. Some vendors further differentiate their offers through bundled capabilities such as binder integration, delivery infrastructure, or private polymorphic stubs. This service model lowers the barrier to entry for less technically sophisticated actors by decoupling operational evasion from the expertise required to design and maintain evasion mechanisms~\cite{DeLaCruz2024, VanWegberg2018}.

On \texttt{exploit.in}, crypters appear not merely as static tools but as continuously maintained operational services whose value depends on recurring stub renewal --- sometimes every 24 to 48 hours --- to remain ahead of antivirus and EDR detection. In this sense, CraaS constitutes a professionalized layer of the cybercrime supply chain, combining technical obfuscation, operational maintenance, and market-based support.

This paper provides a systematic, forum-level empirical analysis of the CraaS ecosystem as observed on \texttt{exploit.in}, one of the most prominent Russian-language cybercrime forums. Our contributions are fourfold: (i)~we characterize the buyer--seller roles and division of labor within the ecosystem; (ii)~we document and compare the dominant economic models and pricing structures; (iii)~we study the co-participation network and show that forum reputation is more closely associated with brokerage and structural influence --- as captured by betweenness centrality and PageRank --- than with raw connectivity alone; and (iv)~we analyze the trust and governance infrastructure --- including escrow services, reputation deposits, and guarantor mechanisms --- that enables stable exchange under conditions of anonymity.

\section{Related Work}
\label{sec:related}

The CraaS ecosystem sits at the intersection of three research bodies: the technical study
of crypters, packers, and obfuscation methods; the empirical analysis of underground
cybercrime markets; and the socio-economic study of trust and governance in illicit
economies. We survey each in turn.

\subsection{Crypters, Packers, and Obfuscation}
Crypters can be understood as an extension of the packer model~\cite{Muralidharan2022} with explicit AV evasion
objectives. 
On the obfuscation side, You and Yim~\cite{You2010} and Brezinski and
Ferens~\cite{Brezinski2023} survey the progression from encrypted to metamorphic malware,
while Alazab et~al.~\cite{Alazab2012} situate these techniques within a cybercrime context.
Barria et~al.~\cite{Barria2016} characterize dead code insertion
(AVFUCKER, DSPLIT) as a complementary static signature disruption strategy. At runtime,
Di Pietro et~al.~\cite{DiPietro2025} show that process injection --- the primary in-memory
execution primitive of crypted payloads --- remains the dominant evasion strategy across
2,667 malware families, with actors continuously adapting variants in response to defenses.

\subsection{Underground Markets and the CraaS Economy}

Huang et~al.~\cite{Huang2017} propose a control-point framework for CraaS ecosystems,
identifying the structural dependencies between service providers and consumers as
actionable disruption opportunities. Van Wegberg et~al.~\cite{VanWegberg2018} provide
empirical evidence that plug-and-play criminal services substantially lower the technical
barrier to entry, while Bhalerao et~al.~\cite{Bhalerao2018} expose the supply chain
dependencies linking seemingly unrelated underground offerings. Akyazi~\cite{Akyazi2021}
further shows, via an eleven-year study of Hack Forums, that CraaS supply and demand remain
structurally stable even following major marketplace takedowns. 
To our knowledge, the only prior work dedicated specifically to CraaS is De La Cruz Alvarado and Pastrana~\cite{DeLaCruz2024},
who analyze vendor social networks, product feature sets, and subscription update dynamics
on \texttt{HackForums}, a clear web forum (see discussions in~Section~\ref{sec:discussion}). 

\subsection{Cybercrime Community Research and Trust}

Hughes et~al.~\cite{Hughes2024} provide a methodological synthesis of cybercrime community
research, cataloguing the epistemic challenges --- ground truth validation, ethical
constraints, platform access --- that confront empirical work in this space, and proposing
best-practice guidelines for future studies. Their framework directly motivates the ethical
and methodological choices we detail in Section~\ref{sec:data}.

\v{S}embera et~al.~\cite{Sembera2021} offer a rare inside view of a specialized
obfuscation service: their analysis of an Android Obfuscation-as-a-Service platform reveals
a niche but profitable operation generating an estimated USD~5,100--61,160 over six months
despite mediocre obfuscation quality. This underscores how low technical quality does not
preclude commercial viability in underground markets, provided pricing and trust mechanisms
are well-calibrated. The companion VB2021 presentation~\cite{Sembera2021b} further
exposes the operational details of how such services manage client relationships and
maintain FUD status.

\subsection{Positioning}
Our work complements and extends prior research on underground crypter markets in three ways. First, it provides a five-year longitudinal analysis of CraaS activity on \texttt{exploit.in}, a restricted-access Russian-language forum, thereby broadening the empirical basis beyond the clear-web setting of \texttt{HackForums}. Second, it offers a finer-grained characterization of buyer and seller roles, pricing models, and capability-transfer logics. Third, it places particular emphasis on the institutional mechanisms that sustain exchange under anonymity, including escrow, deposits, and guarantor-mediated trust.

\section{Dataset and Methodology}
\label{sec:data}

\subsection{Data Collection}
We leverage data collected from \texttt{exploit.in}, a prominent Russian-language cybercrime forum that serves as a hub for threat actors discussing intrusion techniques, malware samples, proof-of-concept exploits, and criminal services. Although \texttt{exploit.in} originates from a predominantly Russian-speaking community, English functions as a lingua franca in commercial threads targeting an international audience. Access required payment of a registration fee (\$200), which provided authenticated access to restricted forum sections. 

Threads were selected by filtering thread titles using the keywords: \textit{crypter}, \textit{cryptor}, \textit{crypt}, \foreignlanguage{russian}{\textit{криптор}}, and \foreignlanguage{russian}{\textit{криптер}}. Only threads created from 2020 onward were considered. In our dataset, 65\% of posts are written in English, compared to 25\% in Russian and 10\% combining both languages. No image content was downloaded automatically; all media collection was performed manually.

Table~\ref{tab:corpus} summarizes the resulting corpus.

\subsection{Ethical considerations}

The data analyzed in this study originate from the underground forum exploit.in, which requires a paid registration to access certain sections of the platform. Access to the forum was obtained through the standard registration procedures of the website, including the payment of the required access fee. This payment was solely intended to obtain observational access to the environment under study and did not involve participation in any illicit activity, transaction, or interaction with forum members.

Given the sensitive nature of the environment studied, particular care was taken to minimize potential risks related to the identification of forum participants. All user pseudonyms were therefore anonymized and replaced with neutral identifiers during the data processing stage. Direct quotations were used sparingly and, when necessary, slightly normalized to reduce the possibility of reverse identification through search engines or forum archives. The analysis presented in this article focuses on aggregate patterns, interaction structures, and economic models rather than on individual actors.

More broadly, this approach follows established methodological practices in the study of online illicit environments, where the objective is to document socio-technical dynamics while limiting the exposure of identifiable individuals and avoiding any interference with the observed ecosystem \cite{DecaryHetu2013, Holt2013, Holt2016, Lusthaus2018, pastrana2018crimebb, VanWegberg2018, Yip2017}.

\begin{table}[h]
\centering
\caption{Corpus Statistics}
\label{tab:corpus}
\resizebox{\columnwidth}{!}{\begin{tabular}{ll}
\toprule
\textbf{Characteristic} & \textbf{Value} \\
\midrule
Threads analyzed   & 491 \\
Total posts        & 2,949 \\
Unique users       & 1,058 \\
Thread creators    & 407 \\
Time window        & Jan.\ 2020 -- Aug.\ 2025 \\
Forum source       & exploit.in \\
Primary languages  & English (65\%), Russian (25\%), Mixed languages (10\%) \\
\bottomrule
\end{tabular}
}
\end{table}

\subsection{Annotation Methodology and Validation}

\begin{figure}[h]
    \centering
    \includegraphics[width=0.5\linewidth]{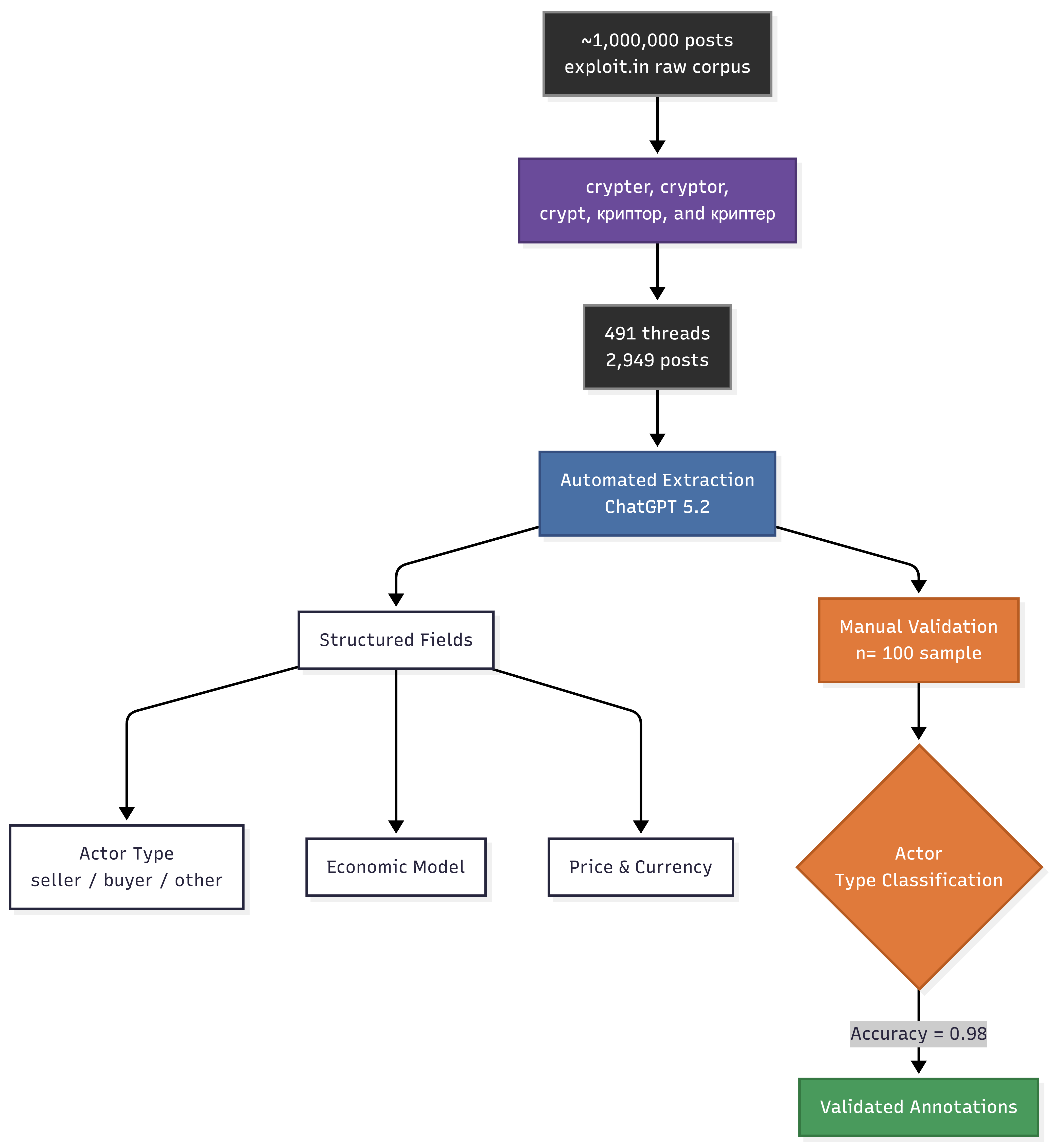}
    \caption{Overview of the annotation pipeline and validation process}
    \label{fig:annotation_validation}
\end{figure}

\begin{figure}[h]
\centering
\includegraphics[width=0.7\linewidth]{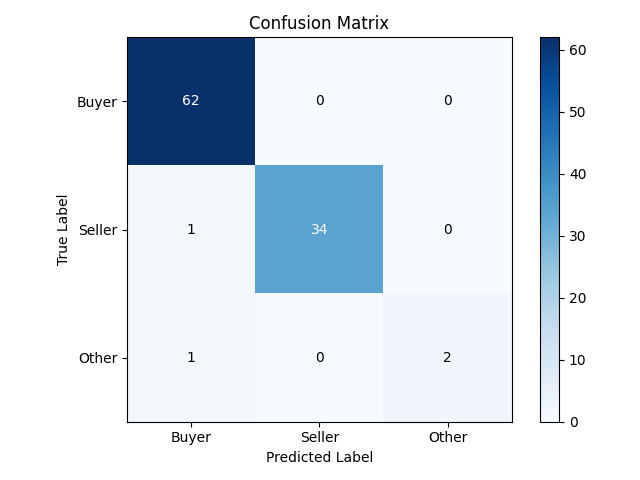}
\caption{Confusion matrix for automated actor type classification (n=100)}
\label{fig:confusion_matrix}
\end{figure}

Each thread was annotated along the following dimensions: actor type (seller, buyer, or other), economic model, price, advertised features, payment currency, and crypter name when explicitly mentioned. The annotation pipeline combined automated extraction with manual validation (Figure~\ref{fig:annotation_validation}). Structured fields were first extracted using ChatGPT 5.2 and subsequently reviewed by human annotators. To assess the reliability of this workflow, we evaluated two key dimensions on a random sample of 100 threads (about 20\% of all threads). For actor-type classification, manual review yielded an accuracy of $0.98$ (Figure~\ref{fig:confusion_matrix}). For economic-model annotation, agreement between model output and independent human annotations reached a Cohen's kappa of $0.96$, indicating almost perfect agreement.

\section{Seller--Buyer Roles in the CraaS Ecosystem}
\label{sec:roles}
In the analyzed dataset, we focus on two primary categories of actors: crypter vendors and buyers. Neither forms a homogeneous group: vendors correspond to several distinct categories with different levels of technical involvement and business models, while buyers exhibit different motivations and operational profiles. Taken together, these observations indicate that the crypter marketplace is structured as a multi-layered ecosystem involving infrastructure providers, professional service operators, resellers, and low-cost vendors, each targeting different segments of the cybercriminal market. The role of influential actors and brokers, whose structural importance is revealed through reputation and centrality analysis, is examined separately in Section~\ref{sec:Network}.

\subsection{A Taxonomy of Crypter Sellers}
\label{sec:taxonomy}

Our analysis identifies five seller archetypes, each characterized by a specific level of technical sophistication, degree of automation, and business model. The following subsections describe each in detail.
\subsubsection{Structured CraaS Operator}

The most professionalized seller archetype operates a full-stack service
comparable to a legitimate Software-as-a-Service : a web panel
for order management, tiered subscription plans, automated payment
processing, regular stub updates, and customer support. These operators
invest in platform infrastructure and forum credibility (evidenced by
moderator-required deposits), and present marketing material including
scan result links, video demonstrations, and changelog updates.

As an illustrative example, one long-standing operator (threads spanning 2022--2025) provides a subscription-based service at \$3,000/month called \texttt{AsmCrypt}~\cite{CybercriminalsUsingNew2023}.

\subsubsection{Independent Software Vendor (ISV)}

Rather than selling a crypting \emph{service}, ISV-type sellers sell the
\emph{tool itself}---a builder application,  or source
code---that enables buyers to crypt their own payloads \emph{independently}
(i.e., without requiring the seller to manually process each payload).

One vendor in this category illustrates the model by selling access to a builder that the buyer operates independently (\emph{``vy kriptuete sami lichno''}---you crypt yourself''). A private stub with auto-generation ensures each build is unique. Subscription: \$200/week, \$600/month, \$1,100/3 months. Notably, the same seller subsequently offered the full source code for \$2,000.

\subsubsection{Telegram Bot Operator}

Closely related to the Structured CraaS Operator but distinguished by the
primacy of the Telegram interface: the client interacts entirely through
a Telegram bot, with near-instant automated delivery and crypto payment
processing. Human support is minimal or absent. This model maximizes
throughput and minimizes operational exposure for the seller.

\textbf{EasyCrypt} exemplifies the low-cost, high-volume end of the Telegram Bot Operator category. Launched in 2022, it advertises itself as one of the most affordable crypters on the market. Pricing starts at \$5 per crypt or \$109/month for unlimited builds, with a paid API tier (\textit{EasyCrypter Plus}, \$25/month). Stub cleaning is performed every four hours. 
It is used by the Dolphin Loader MaaS~\cite{RussianPanda2024}, whose configuration file explicitly
selects \texttt{EasyCrypt} as the crypting backend for LummaC2 payloads.

\subsubsection{Independent Artisan}

Independent Artisan sellers provide mostly manual, bespoke crypting
services with limited automation and limited product packaging. They
typically operate via direct messaging (e.g., forum PM, XMPP/Telegram),
process payloads human-in-the-loop, and emphasize individualized stubs.

One seller (thread title: \emph{"Kriptor ishchet trudoustroystvo"} / "Crypter seeks employment") explicitly advertises handmade stubs and personal availability every weekday. Pricing is \$500--\$1,000/week.

\subsubsection{Fraudulent or Low-Quality Seller}

A significant fraction of sellers in the dataset deliver services that
fail to meet advertised specifications, refuse to provide pre-purchase
tests, disappear after payment, or fabricate review feedback. We label as
fraudulent/low-quality accounts that are repeatedly reported for non-delivery,
refusal to test, or inconsistent claims. This category is not incidental:
experienced forum members estimate that the majority of crypter sellers are
scammers or operate below acceptable quality thresholds.


\subsection{Buyer Taxonomy}
\label{sec:buyer_taxonomy}

\subsubsection{Malware Operator}
\label{sec:buyer_operator}

The \emph{Malware Operator} constitutes the dominant buyer profile in the dataset. These actors maintain active distribution campaigns—typically centered on stealers, RATs, loaders, or botnets—for which crypting serves as a \emph{recurring operational dependency}: each new build must evade current antivirus signatures, and any detection requires rapid re-crypting, often within hours. Monthly budgets range from \$500 to \$3,000, and buyers in this category frequently specify explicit throughput constraints (e.g., \emph{"daily crypt}, \emph{"2–3 times per week"}), as well as target geographies and specific antivirus products aligned with their intended victim population.

\subsubsection{One-Shot Buyer}
\label{sec:buyer_oneshot}

The \emph{One-Shot Buyer} has a single, episodic need: one
crypt for one file, one campaign, or one test.  Crypting is
not a core operational activity but an occasional
requirement.  Budget is correspondingly modest
(\$100--\$1\,000), and buyers in this category typically
require proof-of-work (a scan result or a test crypt) before
committing.  They are disproportionately exposed to
Fraudulent Sellers because they lack the long-term
relationships with trusted vendors that protect more
established operators.

\subsubsection{Tool Acquirer}
\label{sec:buyer_toolacquirer}

The \emph{Tool Acquirer} does not purchase a crypting
service at all; they purchase \emph{crypting capability}---
source code, a builder, or a full business package---so as
to operate independently or to launch their own service.
Their position is structurally at the intersection of buyer
and future seller.  Budgets are the highest in the buyer
taxonomy (\$1\,500--\$20\,000).

\subsubsection{In-House Recruiter}
\label{sec:buyer_recruiter}

The \emph{In-House Recruiter} does not seek a crypting
service on the open market; instead, they seek to
\emph{internalise} crypting capability by hiring a developer
on a salaried, full-time, and exclusive basis.  The
motivations stated in recruitment posts are consistent:
"public" services are too slow, too unstable, or too widely
shared to be operationally reliable, and competitors on the
forum have already been tried and found wanting.  Key
requirements include exclusivity. Monthly salary offers in observed
threads range from \$5\,000 to \$15\,000, with some threads indicating that the first month may be held in escrow as a guarantee.


\subsection{Buyer--Seller Correspondences}
\label{sec:correspondences}

The buyer and seller taxonomies are not independently
structured: each buyer category maps onto one or more seller
categories through a characteristic transactional logic.
Table~\ref{fig:correspondence_matrix} summarises these
mappings, along with the associated pricing ranges and
trust signals observed in the dataset.

\begin{table*}[h]
\centering
\small
\caption{Buyer--seller correspondence matrix}
\label{fig:correspondence_matrix}
\scriptsize
\begin{tabular}{p{2.7cm}p{2.7cm}p{3.4cm}p{2.2cm}p{2.2cm}}
\toprule
\textbf{Buyer} & \textbf{Seller} & \textbf{Logic} & \textbf{Pricing} & \textbf{Trust signals} \\
\midrule
Malware Op.       & Structured CraaS  & Subscription; high vol.; frequent updates   & \$20--\$3\,000/mo  & High (deposit)\\
Malware Op.       & Telegram Bot Op. & Per-build; low friction; high throughput & \$7--\$250/build   & Moderate \\
One-Shot Buyer    & Artisan          & Single crypt; PoW; escrow              &  \$10--\$1k  & Variable \\
One-Shot Buyer    & Fraudulent       & Unintended; info.\ asymmetry           & Any                & None \\
Tool Acquirer     & ISV              & Source code / business pack            & \$1.5k--\$20k (1x) & Variable \\
In-House Recruiter & Artisan         & Internalisation; exclusivity; salary   & \$2k--\$15k/mo & Variable \\
\bottomrule
\end{tabular}
\end{table*}

\subsection{Pricing Model of CraaS}
\label{sec:econ_model}
CraaS markets monetise \emph{evasion as an operational capability} through a small set of recurring revenue models: (i) subscription plans offered by Structured CraaS operators (continuous updates and support), (ii) pay-per-build or pack pricing operated at scale via Telegram bots (low friction, high throughput), (iii) one-shot, human-in-the-loop crypting by artisans (often with proof-of-work and escrow), and (iv) capability transfer via ISVs selling builders, source code, or ``business packages'' enabling buyers to crypt independently or start their own service.
These models reflect a core trade-off between \emph{cost, speed, quality, and exclusivity}: shared services optimise scale but expose buyers to signature overlap, detection and instability, whereas private stubs and in-house hiring internalise the capability at higher cost.
Market governance and information asymmetry strongly shape transactions: forum deposits, escrow, vouches, and trial crypts act as trust mechanisms, while scam prevalence and unverifiable claims remain a persistent failure mode.

\section{Network Structure and Reputation-Centrality Associations on exploit.in} \label{sec:Network}

\begin{figure*}[ht]
    \centering
    \includegraphics[width=\textwidth]{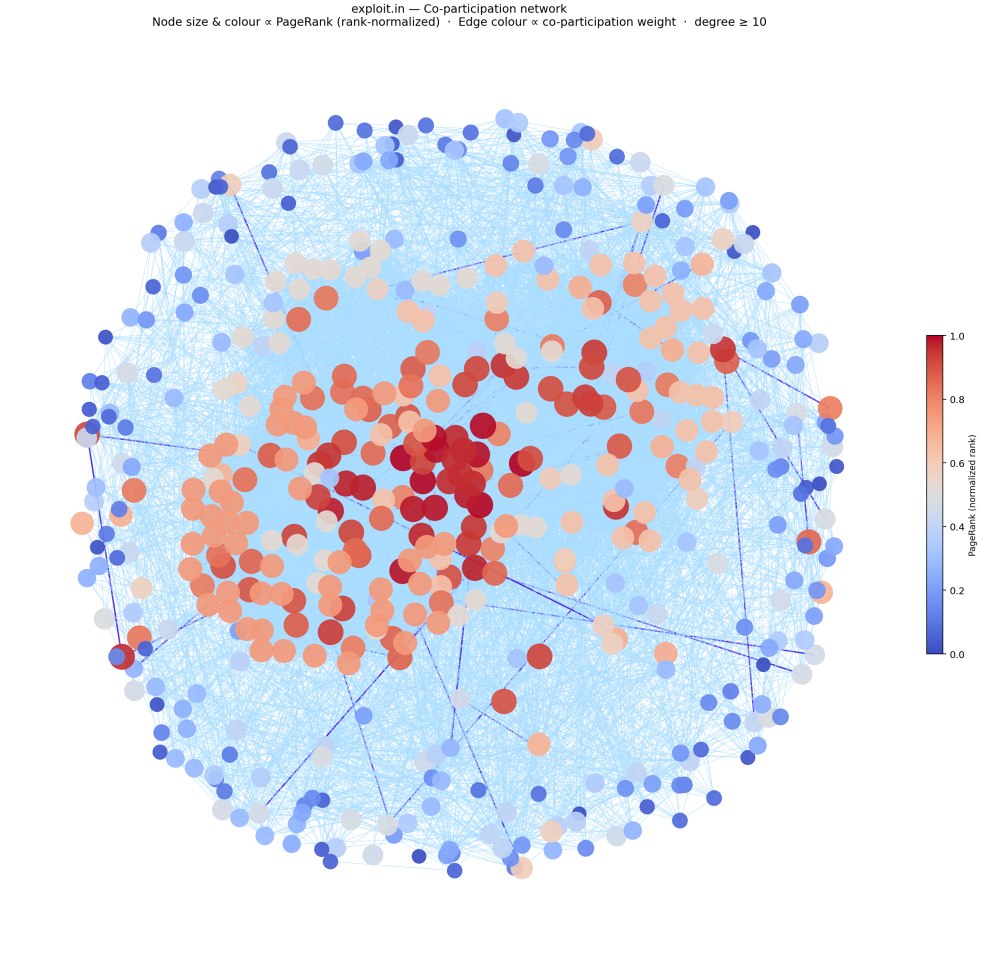}
    \caption{Co-participation network of exploit.in. Node size and colour 
    reflect PageRank (rank-normalized). Edge colour reflects 
    co-participation weight.}
    \label{fig:network}
\end{figure*}

\begin{table}[ht]
    \centering
    \caption{Spearman and Pearson correlations between reputation 
    and centrality measures among thread starters.}
    \label{tab:correlations}
    \begin{tabular}{lrrrr}
        \toprule
        \textbf{Measure} & 
        \textbf{Spearman $\rho$} & \textbf{$p$-value} & 
        \textbf{Pearson $r$} & \textbf{$p$-value} \\
        \midrule
        rep $\sim$ degree      & 0.245 & 0.0003 &  0.106 & 0.1246 \\
        rep $\sim$ betweenness & 0.145 & 0.0350 &  0.501 & 0.0000 \\
        rep $\sim$ pagerank    & 0.204 & 0.0028 &  0.455 & 0.0000 \\
        rep $\sim$ eigenvector & 0.257 & 0.0002 & -0.003 & 0.9672 \\
        \bottomrule
    \end{tabular}
\end{table}

\subsubsection{Network Structure and Centrality-Reputation Associations in the exploit.in Co-participation Graph}
The co-participation graph comprises 903 nodes and 14,549 edges distributed across 36 connected components. Despite this fragmentation, the network is structurally dominated by a single giant component of 815 nodes, accounting for approximately 90\% of all actors. This concentration suggests that most forum interactions unfold within a shared relational space, while the remaining 35 components likely represent peripheral or weakly integrated sub-networks — isolated transaction threads, short-lived exchanges, or actors who never co-participated with the main cluster.
\subsubsection{Methodological caveat: reputation as a partial node attribute}
A critical limitation conditions all subsequent interpretations. Reputation scores are only available for users who initiated at least one thread, a subset that is both smaller and structurally non-random relative to the full forum population. Thread starters are by definition more engaged actors than passive participants, and their structural positions in the network — degree, betweenness, PageRank — are therefore not representative of the forum population as a whole. More critically, the fact that reputation scores are only observable for this non-random subset means that the correlations reported below describe the reputation-centrality relationship within a self-selected group of visible actors, not across all forum participants. This constitutes a form of selection bias: any association found here cannot be extrapolated to the broader network without further evidence.

\subsubsection{Skewness of the reputation distribution and its implications for correlation analysis}
Before interpreting the correlation coefficients, it is important to note that the reputation distribution is highly unbalanced: the vast majority of starters score between 0 and 10, while a single actor stands at an extreme distance from the bulk of the distribution. This outlier profile has direct consequences for the divergence observed between Spearman and Pearson coefficients across all four centrality measures. Given this skew, Spearman's $\rho$ is the more appropriate summary statistic for the general tendency, while Pearson's r should be interpreted as a sensitivity indicator rather than a measure of the population-level linear relationship.

\subsubsection{Degree centrality}
The Spearman correlation between reputation and degree is positive and significant ($\rho$ = 0.245, p < 0.001), suggesting a modest but reliable rank-order tendency for more reputed thread starters to accumulate more co-participation ties. However, the Pearson coefficient is both small and non-significant (r = 0.106, p = 0.125). The pronounced gap between the two statistics points not merely to general non-linearity, but more specifically to the distorting influence of actors who combine extreme reputation with only moderate degree. Their role as independent reviewers can generate very high betweenness and PageRank without necessarily maximizing raw co-participation counts. In short, high reputation does not reliably translate into high degree, even among thread starters.

\subsubsection{Betweenness centrality}
The betweenness result presents the most nuanced profile among the four centrality measures: while the Spearman coefficient is the weakest in rank-order terms ($\rho$ = 0.145, p = 0.035), the Pearson coefficient is the largest across all measures (r = 0.501, p < 0.001). This divergence should not be interpreted as contradictory: the weak Spearman coefficient indicates that brokerage position is not a general property of reputed actors, while the large Pearson coefficient reflects the structural singularity of a handful of actors who simultaneously achieved forum recognition and occupied bridging positions between otherwise disconnected sub-communities.

\subsubsection{PageRank}
The PageRank results follow a similar pattern to betweenness, with a significant Spearman correlation ($\rho$ = 0.204, p = 0.003) and a substantially larger Pearson coefficient (r = 0.455, p < 0.001). The consistency of significant Spearman coefficients across both betweenness and PageRank provides credible evidence that, in rank terms, reputation is more closely associated with positional influence than with the raw number of connections a user accumulates. This is consistent with the qualitative observation that high-reputation actors in this forum tend to occupy roles as reviewers, brokers, or repeat transaction partners rather than as mass participants.

\subsubsection{Eigenvector centrality}
The eigenvector result is the most difficult to interpret. Eigenvector centrality captures embeddedness in a densely connected neighborhood rather than bridging position or raw connectivity, which makes its association with reputation harder to read. The Spearman coefficient is positive and significant ($\rho$ = 0.257, p < 0.001), while the Pearson coefficient is essentially zero and entirely non-significant (r = -0.003, p = 0.967). The Spearman result, while significant, likely captures little more than the fact that a few highly reputed actors also happen to be connected to central neighbors, without this constituting a robust or generalizable tendency.

\subsection{Overall interpretation}
Taken together, the results do not point to a single, uniform relationship between reputation and network position among thread starters. The clearest and most consistent pattern is that reputation is more strongly associated with brokerage and structural influence, as captured by betweenness centrality and PageRank, than with raw connectivity (Degree centrality) or local embeddedness (Eigenvector centrality). This pattern holds even when one accounts for the fact that a small number of extreme actors exert a disproportionate influence on several of the reported coefficients.

Crucially, this finding complements and extends the seller--buyer taxonomy presented in Section~\ref{sec:roles}. While that taxonomy captures the transactional structure of the marketplace, the network analysis reveals a third category of actors --- \emph{trust brokers and influence nodes} --- whose role is not primarily to buy or sell crypting services, but to mediate trust, validate reputations, and connect otherwise disconnected sub-communities. These actors derive their influence not from transaction volume but from their structural position in the network, and their presence suggests that the CraaS ecosystem on \texttt{exploit.in} is sustained not only by commercial exchange but by a broader social infrastructure of trust and intermediation.

This structural interpretation is further supported by the degree distribution across actor types. A striking cross-cutting pattern emerges from the degree statistics: nearly all high-centrality actors are buyers or other participants, not sellers. This asymmetry reflects the structural logic of the marketplace rather than a difference in economic importance: a seller typically remains anchored within their own thread, responding to clients who come to them, while an active buyer traverses multiple threads simultaneously, mechanically accumulating co-participation ties and generating higher centrality scores in the co-participation graph. Centrality in this network is therefore better interpreted as a measure of market mobility than of market power.

\section{Trust and Governance Mechanisms}
\label{sec:trust}

A defining feature of the \texttt{exploit.in} crypter market is its institutionalized
approach to trust management under conditions of anonymity. Rather than eliminating trust,
pseudonymity reshapes it into procedural and economic mechanisms that stabilize exchange.

\subsection{Escrow and Guarantor Services}

Transactions on \texttt{exploit.in} are systematically encouraged---and 
socially enforced---to occur ``strictly through guarantor'' 
(\foreignlanguage{russian}{\emph{Работа строго через гарант}}).  This norm is institutionally enforced by forum moderators, 
who close non-compliant or flagged vendor threads and require 
the payment of a security deposit as a condition of reopening.
 Observed deposit amounts range 
from \$500 to \$1,500, denominated in USD and payable in 
cryptocurrency (Bitcoin or USDT).

The triadic buyer--seller--guarantor structure reduces information 
asymmetry and provides a framework for dispute resolution, 
effectively substituting for contract law in an illicit market where 
formal legal recourse is unavailable. 

Guarantors are not passive  intermediaries: they hold funds in escrow, adjudicate disputes, and 
their involvement is treated as a credibility signal by buyers 
evaluating unfamiliar vendors. 
This norm is enforced not only top-down by moderators, but 
bottom-up by experienced buyers who publicly invoke it and 
signal its violation. One buyer  publicly reproaches 
a vendor for bypassing the forum escrow system: \emph{``there's 
escrow on the forum, that we could use. But you ignored this''} 
(\texttt{exploit.in}).

\subsection{Reputation as Capital}
\texttt{exploit.in}, like many online forums, implements a reputation 
system in which users can assign positive or negative votes to other 
users (\foreignlanguage{russian}{\emph{``зачем вы покупаете крипт у людей без репутации или 
депозита?''}} / \emph{``Why buy from sellers with no reputation?''}, 
\texttt{exploit.in}). A user's reputation is represented by a 
numerical score corresponding to the sum of all votes they have 
received. Reputation on \texttt{exploit.in} operates not merely as 
a social marker but as an economic asset. High-reputation vendors 
command pricing power, and buyers explicitly invoke the absence of 
reputation as grounds for refusing to transact. Newcomers invest in 
signaling strategies: lower introductory pricing 
(\emph{``discount for first 50 clients''}, \texttt{exploit.in}), 
free trials or crypts offered in exchange for public reviews 
(\emph{``FIRST 3 CRYPTS = FREE (for the first to contact us with 
+15 reputation)''}, \texttt{exploit.in}), or publicly demonstrated 
FUD scan results (via avcheck, kleenscan, scanner.to) to overcome 
initial credibility deficits. Refund policies and recrypt guarantees 
--- ranging from 3 to 7 days of post-delivery FUD coverage, with 
free recryption if the stub is burned within the warranty period 
(\emph{``guaranteed 0/26 for 5--7 days; free recrypt if detected''}, 
\texttt{exploit.in}) --- mirror legitimate SaaS accountability 
practices, reinforcing the professionalization of the CraaS model.

\subsection{Structural Implications}

These governance mechanisms collectively enable the division of labor characteristic of
modern cybercrime ecosystems~\cite{Huang2017, VanWegberg2018}. A crypter developer need not
have personal familiarity with downstream malware operators; transactions are mediated
through institutionalized trust, allowing specialization to proceed at scale. Without such
mechanisms, the market would remain fragmented and vulnerable to exit scams, limiting its
professionalization and resilience.

\section{Comparative Discussion: CraaS on HackForums vs. exploit.in}
\label{sec:discussion}
\subsection{Context and Target Audience}
De la Cruz Alvarado and Pastrana Portillo~\cite{DeLaCruz2024} focus their analysis on HackForums,
an English-speaking clear-web forum historically accessible without a paid barrier to entry. Our
corpus, drawn from \texttt{exploit.in}, reflects a markedly different setting: a predominantly
Russian-speaking platform with restricted access and a community largely embedded in Russian- and
Eastern-European cybercrime milieus. This linguistic and access asymmetry is analytically important,
as it suggests partially different user populations and market expectations. Relative to the
HackForums marketplace described in~\cite{DeLaCruz2024}, the \texttt{exploit.in} corpus appears to
capture a more specialized and less publicly accessible segment of the CraaS ecosystem.

\subsection{Technical Sophistication}

One of the most striking divergences concerns the level of technical sophistication of the products
offered. The crypters analyzed on HackForums by~\cite{DeLaCruz2024} are described as standardized
commercial tools equipped with user-friendly graphical interfaces (GUIs), designed for users with
little or no programming expertise. The emphasis is placed on ease of use, file icon or name
customization, and the promise of 24/7 customer support.

On \texttt{exploit.in}, the technical spectrum is far broader. We observe a coexistence of
accessible services (standard .NET stubs priced at \$15--\$60) alongside products of considerable
sophistication: crypters written in Rust with indirect syscalls via inline assembly
(\textit{ADVANCED CRYPT}), bypass of EDR solutions such as CrowdStrike, SentinelOne, and Sophos,
support for DLL/EXE/MSI output formats, Living off the Land (LOLBin) techniques, and advanced
LLVM-based obfuscators featuring opaque predicate generation and Mixed Boolean Arithmetic (MBA)
transformations. Some listings explicitly reference use cases associated with professional red team
engagements, suggesting a partially legitimate or semi-legitimate clientele --- a nuance entirely
absent from the HackForums marketplace described in~\cite{DeLaCruz2024}.

\subsection{Economic Models and Pricing}
\cite{DeLaCruz2024} observe on HackForums a predominance of monthly or annual subscription models,
with prices ranging from approximately \$20 per single crypt to \$550 for a lifetime subscription
with a private stub. Stubs are classified into three categories: standard (shared), pseudo-private
generator, and fully private.

Our corpus reveals a broader pricing range and a finer granularity of economic models. These
include per-unit sales (sometimes only a few dollars for a manual crypt), weekly or monthly
subscriptions, auctions for source code, and high-end offerings reaching several thousand dollars.
Illustrative examples include listings advertising \$8,000--\$12,000/month for an exclusive FUD
runtime stub, or \$25,000 for a mass-infection method. Taken together, this heterogeneity suggests
a market segmented between opportunistic low-budget buyers and more advanced actors seeking durable,
exclusive, or internalized solutions.

\subsection{Trust Mechanisms and Internal Governance}

\cite{DeLaCruz2024} note that HackForums relies primarily on vouches, free trials, and
forum-acquired reputation to establish trust between buyers and sellers, with moderation
remaining relatively inconspicuous in transactional exchanges.

On \texttt{exploit.in}, internal regulation is more formalized and constraining. The use of a
\textit{garant} (an escrow system managed by the forum) is an explicitly encouraged, and at times
enforced, norm. Moderators intervene directly in threads to close listings and require security
deposits (\foreignlanguage{russian}{\textit{депозит}}) ranging from \$1,000 to \$1,800 before any commercial activity may
resume. Disputes between buyers and sellers are publicly arbitrated, and reported scams are openly
discussed. This more institutionalized internal governance infrastructure is indicative of a more
mature marketplace seeking to minimize opportunistic defection.

\subsection{Summary}

In sum, the study by~\cite{DeLaCruz2024} provides a pioneering overview of a relatively accessible
and standardized CaaS market, oriented toward a general English-speaking audience. Our
\texttt{exploit.in} corpus complements this picture by documenting a parallel marketplace that is
more opaque, technically more advanced, more tightly self-regulated, and more deeply integrated
into professional cybercriminal operations. These two markets do not appear to be in direct
competition; rather, they seem to serve different levels of the criminal value chain. This
underscores the need for a comparative, multi-forum approach to fully capture the CaaS ecosystem
in its entirety.

\section{Limitations}
This study has several limitations. Our corpus is restricted to a single forum and to
threads matching crypter-related keywords, which may exclude adjacent services (e.g.,
loaders bundled with crypting) or actors who operate exclusively through private
channels. The LLM-assisted annotation, while validated with high agreement, may
introduce systematic biases on edge cases not captured by our 100-thread sample.
Reputation scores are only available for thread starters, a structurally non-random
subset, which means that the reported centrality--reputation associations cannot be
extrapolated to the full forum population. Finally, our analysis remains observational:
we document advertised prices and claims but cannot verify actual transaction volumes,
service quality, or the operational impact of crypted payloads on downstream campaigns.

\section{Conclusion}

This paper has examined the Crypter-as-a-Service ecosystem on \texttt{exploit.in}, a major Russian-language cybercrime forum, through a longitudinal analysis of threads published between 2020 and 2025. Our results show that crypters are not merely marketed as standalone tools, but as operational services whose value depends on continuous maintenance, evasion performance, and trust-based delivery.

Beyond the technical dimension of obfuscation, the market appears structured by a differentiated division of labor. We identify multiple seller and buyer profiles, ranging from structured service operators and Telegram-based low-cost providers to one-shot buyers, tool acquirers, and in-house recruiters. This heterogeneity indicates that CraaS is better understood as a service ecosystem than as a single product category.

We further show that exchange is sustained by layered trust and governance mechanisms, including escrow, moderator intervention, security deposits, and reputation-based signaling. Network analysis complements this view by suggesting the presence of highly central actors whose role extends beyond direct buying and selling, and includes brokerage, validation, and trust mediation.

Finally, our comparison with prior work on \texttt{HackForums} suggests that the \texttt{exploit.in} marketplace reflects a less accessible, more strongly governed, and in several respects more professionalized CraaS configuration. More broadly, the findings support the view that crypter markets occupy an important intermediary position in the malware economy, connecting technical obfuscation capabilities with scalable criminal service provision.

Future work could extend this analysis in three directions: cross-forum comparison, finer-grained longitudinal tracking of seller trajectories and pricing dynamics, and closer integration between forum observations and malware-centric technical analysis.

\section{Acknowledgement}

This work has benefited from the European Union’s Horizon
Europe research and innovation programme ENSEMBLE under
grant agreement No 101168360; and from the French National Research Agency (ANR) under the Plan France 2030, bearing the reference ANR-22-PECY-0007.
We also thank the Cambridge Cybercrime Centre for providing access to the CrimeBB dataset \cite{pastrana2018crimebb}, which contains scraped threads from Hack Forums.

\bibliographystyle{IEEEtran}
\bibliography{wacco}

\end{document}